\documentclass[journal]{IEEEtran}
\usepackage{amsmath,amssymb,bm}  
\usepackage{cases}
\usepackage[overload]{empheq}
\usepackage{graphics}
\usepackage[export]{adjustbox}
\usepackage{epsfig}
\usepackage{booktabs}
\usepackage{float}
\usepackage{cite}
\usepackage{url}
\usepackage{subfigure}
\usepackage{multirow}
\usepackage{siunitx}
\usepackage{eurosym}
\usepackage{enumerate}
\usepackage{tabularx}
\usepackage[super]{nth}
\usepackage{caption}
\usepackage{graphicx}
\usepackage[font={sf,scriptsize},           
            labelfont={bf,color=black},
            caption=false]                  
           {subfig} 
\usepackage{setspace} 
\usepackage[linesnumbered,ruled,vlined]{algorithm2e}

\newcommand*\xbar[1]{%
  \hbox{%
    \vbox{%
      \hrule height 0.5pt 
      \kern0.5ex%
      \hbox{%
        \kern-0.1em%
        \ensuremath{#1}%
        \kern-0.1em%
      }%
    }%
  }%
}

\IEEEoverridecommandlockouts
\title{\LARGE \bfseries Semi-automated transmission control for motorcycle gearshift: design, data-driven tuning and experimental validation}

\author{Edoardo Catenaro, Giulio Panzani, Davide Sette and Sergio M. Savaresi
	\thanks{E. Catenaro,  G. Panzani and S. M. Savaresi are with the Department of Electronics, Information and Bioengineering, Politecnico di Milano, Milan, Italy. Corresponding author: Giulio Panzani.}
	\thanks{D. Sette is with Ducati Motor Holding, Bologna, Italy.}
}

\begin{document}
\maketitle

\begin{abstract}
	This brief addresses the gearshifting problem for Semi-Automated Manual Transmissions (S-AMT) in powered two-wheelers, a powertrain setup that allows fast and smooth gear shifts with minimal modifications to the traditional manual powertrain layout. We show that with a proper synchronization between the electronic clutch and engine torque, excellent gearshift performance can be obtained, but requires precise parameters calibration. We thus propose the use of a data-driven approach with Constrained Bayesian Optimization to optimize control parameters. The procedure’s effectiveness is demonstrated on a real vehicle, assessing performance in terms of optimality, convergence rate, and repeatability.
\end{abstract}


\section{Introduction}
The adoption of automatic transmissions for two-wheeled vehicles is far less common than in four-wheeled vehicles. Apart from simple Continuously Variable Transmissions (CVT) solutions, mainly used in lightweight and mass-market city vehicles, the high-performance two-wheelers automatic transmission market is dominated by Dual Clutch Transmission (DCT) -- a fully automated powertrain layout, that uses two clutches to achieve fast and smooth gearshifts. When the adoption of a such sophisticated powertrain layout is not feasible, the so-called \textit{Quick-Shift} gearshifting control technology is an option, which relies solely on engine torque manipulation to achieve rapid gearshifts but at the cost of significant mechanical stress on the transmission. 

With the diffusion of ride-by-wire systems, the Automated Manual Transmission (AMT) architecture, rooted on the traditional manual powertrain layout, has gained renewed interest. In particular, a simplified version of AMT, the Semi-Automated Manual Transmission (S-AMT), features electronic control over engine torque and clutch while the driver still operates the gearshift pedal manually, avoiding the complex powertrain redesign needed to accommodate for an automatic pedal actuation. As such, the S-AMT can be seen as an enhancement of the \textit{Quick-Shift} gearshifting control technology.

The scientific literature on automatic transmissions in two-wheelers is limited. DCT layouts face similar challenges as those in four-wheeled vehicles, resulting in only a few studies focused on two-wheelers, see \textit{e.g} \cite{watanabe2011development}. Some research explores innovative powertrain layouts based on planetary gears \cite{sheu2002dual,chung2014energy} and the remaining literature primarily addresses AMT powertrain solutions \cite{ahmed2008electronic,2014Giani,neghinua2017robotized,shetkar2022implementation}. To the best of the authors' knowledge, to date, no scientific contributions have addressed S-AMT motorcycle transmissions, which is the focus of this work. The main challenge for Semi-Automated Manual Transmission is dealing with the dual manual-automatic nature of the gearshifting process. Indeed, the manual part of the gearshift prevents keeping under control the entire gearshift dynamics and, additionally, introduces more variability in the process. 

In this work, we first propose in Section \ref{sec:system_description} a control strategy for both engine torque and the clutch-by-wire system, highlighting its benefits over the traditional \textit{Quick-Shift} solution. Precise parameters calibration is essential to achieve the proper synchronization among the actuators and the manual operations. Gearshifting spans multiple physical domains -- electrical, hydraulic, and mechanical -- which complicates the accurate modeling of powertrain components. Relying solely on a simulation or model-based approach to calibrate the gearshift controller is, in fact, impractical. Being aware of these difficulties and aiming to propose an end-to-end solution, in Section \ref{sec:data_driven_optimization_framework} we also detail the process for tuning the control parameters, which is based on Constrained Bayesian Optimization, an iterative data-driven global optimization technique. Section \ref{sec:experimental_results} presents the performance of the proposed tuning strategy and control system, supported by an extensive experimental campaign conducted on a prototype vehicle.

\section{Semi-automatic electronic transmission}\label{sec:system_description}
\subsection{System layout and experimental setup}
The Semi-Automated Motorcycle Transmission (S-AMT) tested in this work is implemented on a prototype \textit{Ducati Diavel 1260S} motorcycle, shown in Figure \ref{fig:control_architecture}. The key additional components of the prototype are:
\begin{itemize}
	\item a Clutch-By-Wire (CBW) actuator that operates the clutch, featuring a low-level control system for clutch pressure reference tracking; details of this control loop, implemented per state-of-the-art electro-hydraulic actuator guidelines, are beyond the paper's scope.
	\item a gear foot pedal (also called \textit{Quick-Shift} pedal), equipped with an electronic switch that detects the driver's shift intent, and generates a \textit{Quick-Shift} ($qs$) signal;
	\item a real-time prototyping Electronic Control Unit responsible for implementing the control strategies. 
\end{itemize}

\begin{figure}[!h]
	\centering
	\includegraphics[width=\linewidth, trim={0cm 1.8cm 0cm 0cm}, clip]{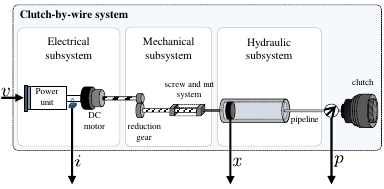}
	\caption{ \small{Pictorial representation of the electro-hydraulic clutch-by-wire (CBW) system.} }
	\label{fig:actuator}
\end{figure}

The CBW system is schematically illustrated in Figure \ref{fig:actuator}. In summary, the traditional hydraulic circuit that operates the clutch has been disconnected from its manual controls and replaced with an electro-mechanical assembly, which includes an electric motor and a transmission. The DC motor, powered by a control unit, pushes the master cylinder of the hydraulic circuit, building the pressure needed to open the clutch.

\subsection{Control strategy}
During a S-AMT gearshift, three actuators must be properly synchronized: the gear selector (pedal lever), the engine torque, and the CBW pressure. The pedal lever is operated by the driver, who pulls it for upshifts or pushes it for downshifts using their foot. Engine torque and CBW pressure are electronically controlled by adjusting the corresponding reference signals, in order to temporarily stop of the engine's power flow through the transmission, enabling the mechanical engagement of the new gear and preventing engine over/under speed.

Compared to traditional ATM controls, which typically operates the actuators in a closed-loop fashion targeting a certain transmission output torque and clutch slippage \cite{2006Glielmo,pisaturo2014multiple,wang2018position,guercioni2019gearshift}, motorcycle S-AMT transmissions require very fast actuator operations. Indeed, manual gearshifts are already rapid maneuvers (see \cite{2011Giani}), and the coexistence of partially manual operations (i.e., gear pedal push/pull) in the S-AMT gearshift process imposes strict time constraints, as the driver ultimately dictates the gearshifting dynamics. As such, closed-loop-based operations of automated AMTs are not a viable control strategy option, and open-loop dominated approaches must be preferred \cite{gao2015optimal}. 

The proposed control strategy is inspired by the open-loop control strategy presented in \cite{2014Giani}, for a sport motorcycle ATM. The fully automated nature of the powertrain considered in the referenced work, however, makes the setpoint definition relatively straightforward, with the main concerns focusing on the calibration of the low-level controllers for the CBW system. In the case of the S-AMT transmission, due to the presence of a manual component in the process, achieving proper synchronization of the various shifting phases is not easily guaranteed, making it challenging to define setpoint profiles for the various actuators. As a result, the first challenge we need to address is selecting a (preferably small) set of key parameters that enable a high-performing gearshift. The second challenge involves proposing a practical and effective method for calibrating these parameters, which motivates the importance we place on data-driven calibration.

\begin{figure*}[!h]
	\centering
	\includegraphics[width=0.8\textwidth]{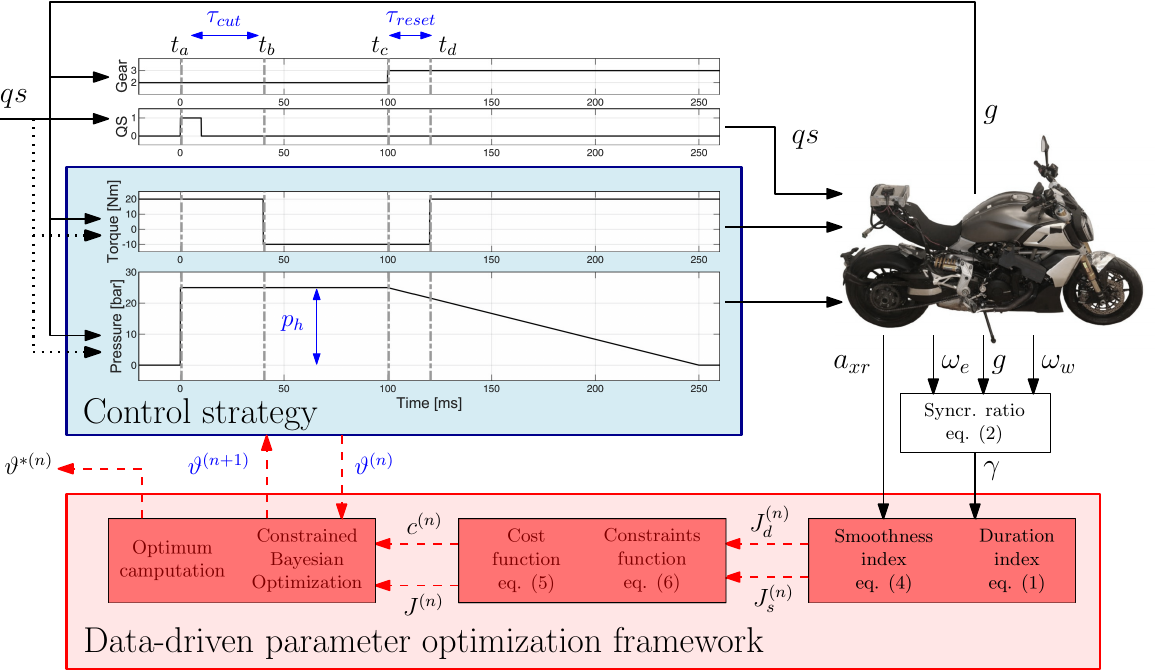}
	\caption{Control strategy and data-driven parameter optimization framework representation.}
	\label{fig:control_architecture}
\end{figure*}

The overview of the proposed control strategy is depicted in Figure \ref{fig:control_architecture}, for a typical upshift maneuver. The gearshift is initiated at time $t_a$ by the pedal action of the driver (\textit{QS} signal in the upper plot). In response, the CBW pressure is immediately increased so to open the clutch and the engine torque is reduced up to its minimum value at time $t_b = t_a + \tau_{cut}$. In this way, both the powerflow interruption and the engine overspeed objectives are achieved. The different timings for manipulating engine and clutch references are due to their distinct actuation bandwidths: indeed, the CBW pressure loop takes about 100 ms to reach maximum value while the ECU responds rapidly to torque changes, motivating the adoption of a delayed request. As soon as the new gear is inserted at time $t_c$, the CBW pressure is gradually reduced and the engine torque is restored to the original value at time $t_d = t_c + \tau_{reset}$.

The control strategy is fully defined by a set of parameters, whose precise calibration is essential. In order to keep the calibration phase as simple as possible, only a minimal subset of parameters has been designated as tunable, while the remaining ones have been fixed at constant values. In particular, the tunable parameters are $p_{high}$ and $\tau_{cut}, \tau_{reset}$. The first parameter correlates with the CBW opening level: while a too small value would keep the clutch closed, waving the potential CBW benefits, a too large value could result in an excessively long gearshift. The other two parameters are associated with changes in torque requests and, as previously discussed, are intended to properly synchronize engine operations with those of the clutch. The lower engine torque request and the CBW closing time have been set to constant values. The engine torque cut request is set to the minimum possible value to ensure that no engine power is delivered during the gearshift. The CBW closing time $T_{close}$ regulates the clutch closing after the gearshift has occurred. A preliminary experimental campaign has shown that the value of $T_{close}$ has a secondary effects on overall gearshift performance: a value of $T_{close} = 200ms$ has been found to strike the best compromise between gearshift speed and clutch closing smoothness.

We highlight how the proposed control strategy is generally valid for any of gearshift. The only consideration to keep in mind is that if the driving torque is negative (as frequently occurs during downshifts), instead of executing a torque cut, the torque request should be positive. This ensures the transition through the engine null power condition, facilitating the engagement of the new gear.

\subsection{Gearshift quality assessment}\label{sec:gear_shift_quality_assessement}
For effective and objective automatic parameter tuning a quantitative assessment of gearshift quality is essential. It is widely recognized, see \textit{e.g.} \cite{guercioni2019gearshift} that gearshift quality is primarily related to two aspects: its \textit{duration} and its \textit{smoothness}. The trade-off between these objectives is evident: shorter shifts can easily result in abrupt oscillations in the driveline due to rapid variations in engine RPM, causing mechanical stress for the driveline and resulting in riding discomfort. Conversely, smoothing engine speed transitions can lead to excessively long gearshift durations.

The employed \textit{duration} index:
\begin{equation}
	J_{d} = t_{e} -  t_{a}
	\label{eq:J_d}
\end{equation}
measures the time elapsed between the driver’s request for a gearshift ($t_{a}$) and the completion of the gearshift ($t_{e}$). We define the gearshift completion time as the moment when the engine and rear wheel rotational speeds are \textit{syncronized} \cite{2011Giani}. To determine when this condition occurs we use the normalized transmission ratio $\gamma$:
\begin{equation}
	\gamma = \frac{\omega_{w}\gamma_{g}(g)}{\omega_e},
	\label{eq:gamma}
\end{equation}
which compares the engine speed $\omega_e$ with the rear wheel angular speed, referenced to the engine according to the selected gear transmission ratio $\gamma_{g}(g)$. Ideally, synchronization occurs when $\gamma = 1$. To account for measurement noise, the final time of the gearshift maneuver coincides with the first instant when $\gamma$ enters the band $1 \pm 0.03$ and remains within it.
\begin{equation}
	t_{e} = {t} : \Bigg\{ | 1 - \gamma(\tilde{t})| \leq 0.03, \; \forall  t > \tilde{t}, \; \tilde{t} \in \mathbb{R}, \; t \in \mathbb{R} \Bigg\}.
	\label{eq:t_e}
\end{equation}

The evaluation of the gearshift smoothness can be either \textit{driver}-oriented or \textit{driveline}-oriented. The former approach focuses on analyzing vehicle longitudinal dynamics as a proxy for the driver's longitudinal motion, by monitoring vehicle speed and acceleration, see \textit{e.g.} \cite{2011Giani}. The latter approach monitors oscillations in powertrain signals (see, e.g., \cite{2020Kawakami}), tightly related to its mechanical stress. In this paper, we establish a meaningful compromise by defining the \textit{smoothness} index $J_{s}$ as the root-mean-square (RMS) value of the high-pass rear wheel's longitudinal acceleration $a_{xr}$ computed over the time window $t_a \leq t \leq t_s$, which has been selected to be sufficiently large ($t_s - t_a = 800$ms) to surely capture the entire gearshift process.
\begin{equation}
	J_{s} =  \sqrt{\frac{T}{t_{s} - t_a} \sum_{i=t_a}^{t_{s}} a_{xr}(i)^2 }.
	\label{eq:J_s}
\end{equation}
In equation \eqref{eq:J_s}, the symbol $T$ indicates the signal sampling time.

\subsection{Comparison with \textit{QS} strategy}
We now present a comparison of two upshifts from 2nd to 3rd gear, which differ in parameter values. The first gearshift is labeled with the acronym \textit{QS}. For this maneuver, the CBW remains closed by setting $p_{high} = 0$ and only the engine torque is utilized. This method of operating the powertrain during a gearshift is known as \textit{Quick-Shift} (\textit{QS}) technology, and is commonly employed by motorcycle manufacturers because it does not require any modifications to the classic manual powertrain layout. It is designed to enhance vehicle performance by enabling very fast gearshifts and does not require any clutch manipulation by the rider. The \textit{QS} strategy, implemented and optimized by the vehicle manufacturer, uses the parameter values reported in Table \ref{tab:QSparam}. It serves as a state-of-the-art benchmark for the second control strategy.
\begin{table}[!h]
	\centering
	\begin{tabular}{@{} p{1.2in} p{0.8in} l @{}}
		\toprule
		\textbf{Description} & \textbf{Symbol} & \textbf{Opt. value} \\ 
		\midrule
		Cut torque delay & $\tau_{cut}$ & 28~ms  \\
		Reset torque delay & $\tau_{reset}$ & 38~ms  \\
		\bottomrule
	\end{tabular}
	\caption{ \small{Optimal parameter values for the \textit{QS} control strategy.} }
	\label{tab:QSparam}
\end{table}
The second approach to gearshifting -- called \textit{QS-CBW} -- uses the additional freedom of clutch control to achieve greater decoupling between the engine and powertrain during gear shifts. Indeed, relying solely on engine torque results in very rapid shifts, but the sudden changes in engine speed cause mechanical shocks to the powertrain, which can eventually lead to transmission damage and result in overall uncomfortable gearshifts.
\begin{figure}[!h]
	\centering
	\includegraphics[width=\columnwidth]{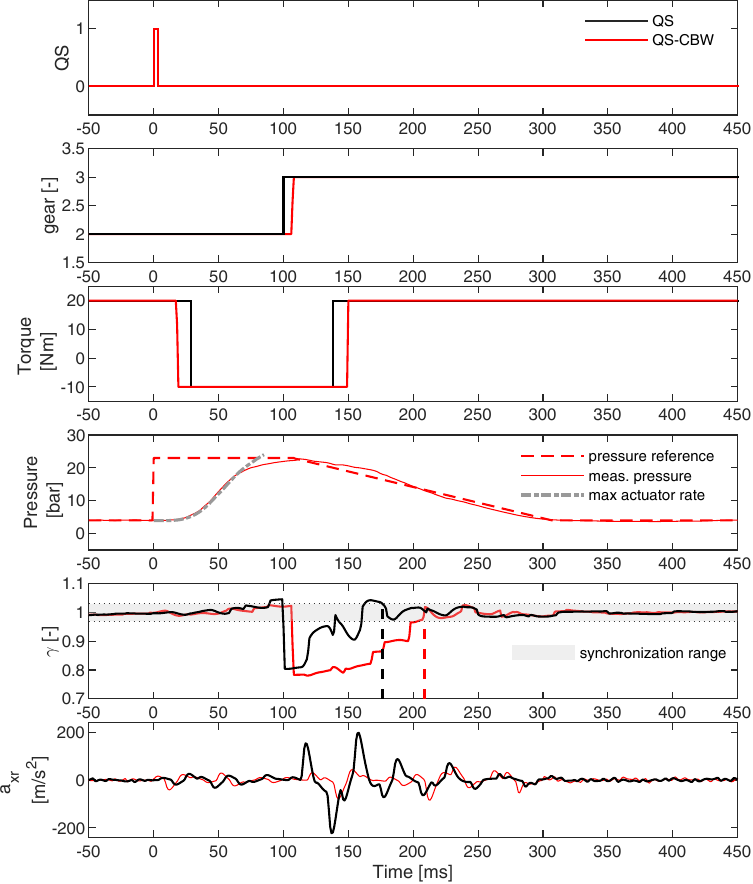}
	\caption{ \small{Time domain response of a QS-based logic (black) compared with respect to the optimized mixed QS-CBW (red) solution.} }
	\label{fig:time_dom_comp}
\end{figure}

The inspection of Figure \ref{fig:time_dom_comp} supports the previous considerations. The \textit{QS}-based logic shortens the duration of the maneuver (see the dashed vertical lines) but, as expected, pays the price of high accelerations peaks when the new gear is engaged, followed by prolonged powertrain oscillations. In contrast, the \textit{QS-CBW} strategy takes advantage of the CBW's actuator action: eventually, the gearshift completion time $t_e$ is slightly worse compared to the \textit{QS}-based maneuver but it offers significant advantages in terms of reduced mechanical stress and gearshift smoothness.
\begin{table}[!h]
	\centering
	\begin{tabular}{@{} p{1.2in} p{0.8in} l @{}}
		\toprule
		\textbf{Description} & \textbf{Symbol} & \textbf{Opt. value} \\ 
		\midrule
		Step amplitude & $p_{high}$ & 23.1~bar  \\
		Cut torque delay & $\tau_{cut}$ & 19~ms  \\
		Reset torque delay & $\tau_{reset}$ & 44~ms  \\
		\bottomrule
	\end{tabular}
	\caption{ \small{Optimal parameter values for the \textit{QS-CBW} control strategy.} }
	\label{tab:QSCBWparam}
\end{table}

\section{Data-driven gearshift parameters optimization}\label{sec:data_driven_optimization_framework}
Given the multi-domain physics involved in the gearshifting process --  electrical, hydraulics, mechanical -- the challenge of properly tuning the control strategy parameters is well-known to the field experts. The rising interest in data-driven techniques and machine learning, pushed the scientific literature to focus on this problem. Two main types of contributions emerge: first, iterative learning policies are applied to refine the performance of closed-loop controllers \cite{lu2014data,yahagi2020slip,mishra2021automated,beaudoin2022improving,dong2023gearshift}; second, iterative data-driven global optimization techniques are applied to fine-tune controller parameters \cite{2011Pinte,2012Vaerenbergh,tanelli2011transmission}. 

Considering the open-loop nature of the discussed \textit{QS-CBW} control strategy, we follow the latter approach type. In particular, the \textit{QS-CBW} parameter tuning procedure is based on the solution of an optimization problem, in the form:
\begin{equation*}
	\begin{aligned}
		\vartheta^* = \operatorname*{argmin}_{\vartheta \in \Theta} \quad & J(\theta)\\
		\textrm{s.t.} \quad & c_s(\vartheta) \leq \overline{c}_s, \; s \in \{ 1,...,S \} \\
	\end{aligned}
	\label{eq:opt_prob}
\end{equation*}
where:
\begin{itemize}
	\item $\vartheta \in \Theta$ are the optimization variables
	\item $J(\vartheta): \Theta \rightarrow \mathbb{R}$ is the cost function
	\item $c_s(\vartheta): \Theta \rightarrow \mathbb{R}, \; s \in \{ 1,...,S \}$ are the constraint functions and $\bar{c}_s$ their boundaries
\end{itemize}

The optimization parameters are $\vartheta = \left\{ p_{high}, \tau_{cut}, \tau_{reset} \right\}$, namely the maximum pressure reference, the torque reduction delay, and the torque restoration delay.

The cost function $J$ is selected to assess the quality of the gearshift maneuver. Although gearshift smoothness is the main concern, shorter maneuver should be favored. Accordingly, the cost function is defined as a convex combination, through a parameter $\lambda \in (0, 1)$, of both smoothness and duration indexes:
\begin{equation}
	J(\vartheta) =  \lambda \left( \frac{J_{s} - \mu_{J_{s}}^{\text{\tiny{\textit{QS}}}}}{\sigma_{J_{s}}^{\text{\tiny{\textit{QS}}}}} \right) + (1-\lambda) \left( \frac{J_{d} - \mu_{J_{d}}^{\text{\tiny{\textit{QS}}}}}{\sigma_{J_{d}}^{\text{\tiny{\textit{QS}}}}} \right).
	\label{eq:J_tot}
\end{equation}
The costs $J_{s}$ and $J_{d}$ are normalized using the mean ($\mu$) and standard deviation ($\sigma$) of smoothness and duration indices, computed from a large dataset of \textit{QS} maneuver that serves as a performance reference. In this way, a straightforward interpretation of the cost function values arises: a value close to zero indicates gearshift performance similar to QS ones, a negative value indicates improved gearshift quality, and a positive value indicates worsening performance.

Box constraints on parameter values are easily enforced in the problem. While gearshift duration can be penalized by lowering $\lambda$, there is no guarantee on the maximum maneuver duration, which is a crucial requirement. Therefore, we introduce an additional constraint to forbid parameter values that lead to unacceptable gearshifts durations. The overall problem constraints are then defined as:
\begin{equation}
	\begin{array}{lrl}
		c_u: & \theta        & \leq \bar{\theta}_u  \\
		c_l: & -\theta       & \leq \bar{\theta}_l \\
		c_t: & J_{d}(\theta) & \leq \bar{c}_t
	\end{array}
	\label{eq:c}
\end{equation}
where $\bar{\theta}_u$ and $\bar{\theta}_l$ are the upper and lower parameter search bounds, while $\bar{c}_t$ is the maximum admissible gearshift duration.

Since model-based optimization is not viable, our approach uses measured data to directly guide the optimization process rather than identifying a system model. We leverage the \textit{Constrained Bayesian Optimization} (CBO) algorithm, a model-free and gradient-free sequential optimization strategy, which can also incorporate constraints into the optimization problem. The most relevant features of the CBO algorithm are summarized; for further details on CBO, interested readers may refer to \cite{2014Gardner}. The CBO algorithm employs surrogate statistical models to estimate both the unknown cost function $J(\vartheta)$ and the constraint functions $c_i(\vartheta)$. These functions are modelled as \textit{Gaussian Processes} (GP), \textit{i.e.} scalar Gaussian distributions characterized by mean $\mu$ and covariance $\sigma$ parameters which vary according to $\vartheta$:
\begin{equation}
	\begin{array}{rcl}
		J({\vartheta})  & \approx & \hat{J}(\vartheta) \sim  \mathcal{N} \left( {\mu}_{J}({\vartheta}), {\sigma}_{J}({\vartheta}) \right)  \\
		c_i({\vartheta})& \approx & \hat{c}_i(\vartheta) \sim \mathcal{N} \left( {\mu}_{c_i}({\vartheta}),{\sigma}_{c_i}({\vartheta}) \right).
	\end{array}
	\label{eq:GP}
\end{equation}
At each iteration of the algorithm, a gearshift is executed using specific values for the parameters $\vartheta^{(n)}$. The algorithm's initialization phase aims to gain a preliminary understanding of the objective and constraint functions by conducting the first $N_r$ experiments, where the parameters $\vartheta^{(n)}$ are randomly selected within the search range. Then, in each subsequent iteration, the following operations are performed (see also Figure \ref{fig:control_architecture}):
\begin{enumerate}
	\item The values of the smoothness index $J_s^{(n)}$, the duration index $J_d^{(n)}$ as well as the corresponding cost function ${J}^{(n)}$ and constraint functions ${c}_i^{(n)}$ are computed.
	\item The approximations of the surrogate cost function and constraint functions are updated with the newly gathered data and a new candidate test point $\vartheta^{(n+1)}$ is computed, as the one that maximizes the \textit{constrained expected improvement} \cite{2014Gardner}. This selection criterion promotes the parameter values that offer the greatest cost improvement (w.r.t to the current optimum) with the highest probability. The probability of constraint satisfaction for the candidate sample is also considered, based on the surrogate models of the constraints.
	\item The optimal parameter value $\vartheta^{*(n)}$ is computed, being the \textit{best feasible visited sample} considering all the recorded values of ${J}^{(n)}$ and ${c}_i^{(n)}$.
\end{enumerate}
The algorithm's stopping criterion ensures the optimization process concludes in a timely manner, and it is based on a maximum number of iterations $N$.

\section{Experimental results}\label{sec:experimental_results}
For the sake of clarity and conciseness, we focus the analyses of the collected experimental data on a specific gearshift, namely the $2 \rightarrow 3$ upshift. Indeed, the most critical shifts are those involving lower gear ratios, where the engine inertia is comparable with the vehicle one and engine speed changes have the highest impact on the powertrain. We chose not to focus gear transitions involving the $1$st gear because they necessitates a transition through the neutral position, resulting in greater variability in the experimental results and unnecessarily complicating the analysis. However, the presented results remain representative and applicable to any other gearshift. Table \ref{tab:ParamExpRes} reports the CBO algorithm parameters employed for the experimental campaigns.
\begin{table}[!h]
	\centering
	\begin{tabularx}{\columnwidth}{@{} p{0.8cm} X p{4cm} @{}}
		\toprule
		\textbf{Param.} & \textbf{Value} & \textbf{Description} \\ 
		\midrule
		$\bar{\theta}_l$ 	& $\left[10 \, \text{bar}, 0 \, \text{ms}, 0 \, \text{ms}\right]$ & Lower parameter bounds \\ 
		$\bar{\theta}_u$ 	& $\left[30 \, \text{bar}, 100 \, \text{ms}, 100 \, \text{ms}\right]$ & Upper parameter bounds \\ 
		$\lambda$ 			& 0.8   		& Performance weight \\ 
		$N$       			& 50    		& Experiments budget \\ 
		$N_r$       		& 4    			& Initial random samples \\ 
		$\bar{c}_t$ 		& 0.5 \, \text{s} & Duration constraint \\ 
		\bottomrule
	\end{tabularx}
	\caption{\small{CBO algorithm parameters during the experimental campaign.}}
	\label{tab:ParamExpRes}
\end{table}

\subsection{Optimization procedure analysis}
\begin{figure}[H]
	\centering
	\includegraphics[width=\columnwidth]{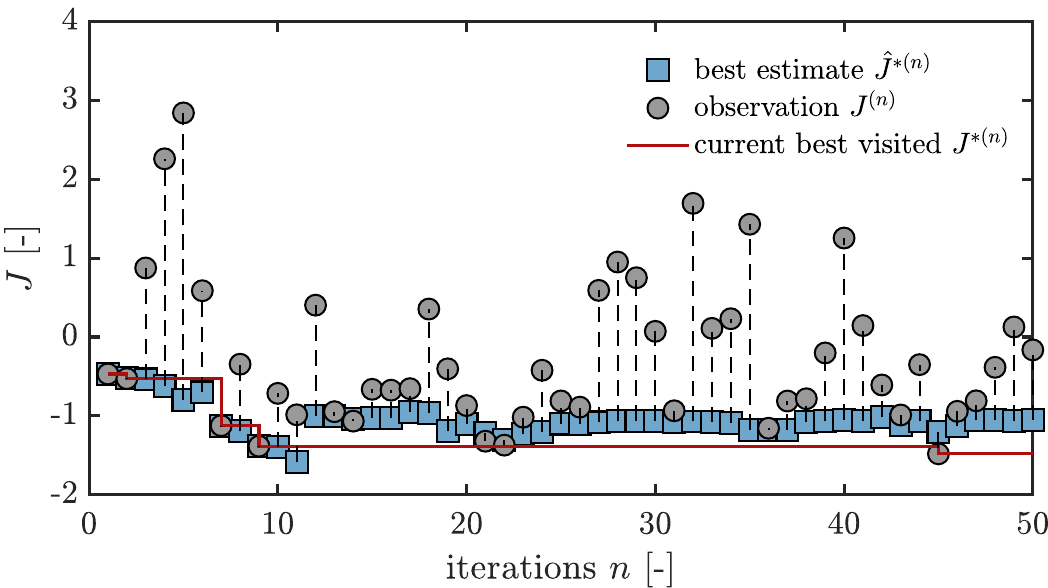}
	\caption{ \small{Measured and estimated cost function values during an automatic optimization campaign.} }
	\label{fig:algorithm_iter}
\end{figure}

First, we present the behavior of a single CBO run. Figure \ref{fig:algorithm_iter} illustrates the evolution of the sampled observations of the cost function ${J}^{(n)}$ and the optimal cost of the \textit{estimated} surrogate model ${\hat{J}}^{*(n)}$. The cost value of the best \textit{visited} feasible sample ${J}^{*(n)}$ is also reported.
Two distinct phases of the algorithm can be observed. During the first phase -- lasting approximately until iteration 25 -- the algorithm actively searches for the minimum of the cost function: the observed costs ${J}^{(n)}$ are very close to the estimated optimum  ${\hat{J}}^{*(n)}$, and the algorithm samples around the optimal parameter region to filter out experimental variability. In the second phase, after the location of the global minimum has been identified, it explores previously unsampled parameter regions to reinforce its conclusions. Accordingly, the sampled costs yield significantly higher values, further confirming the location of the estimated optimum.

\begin{figure*}
	\centering
	\includegraphics[width=\textwidth]{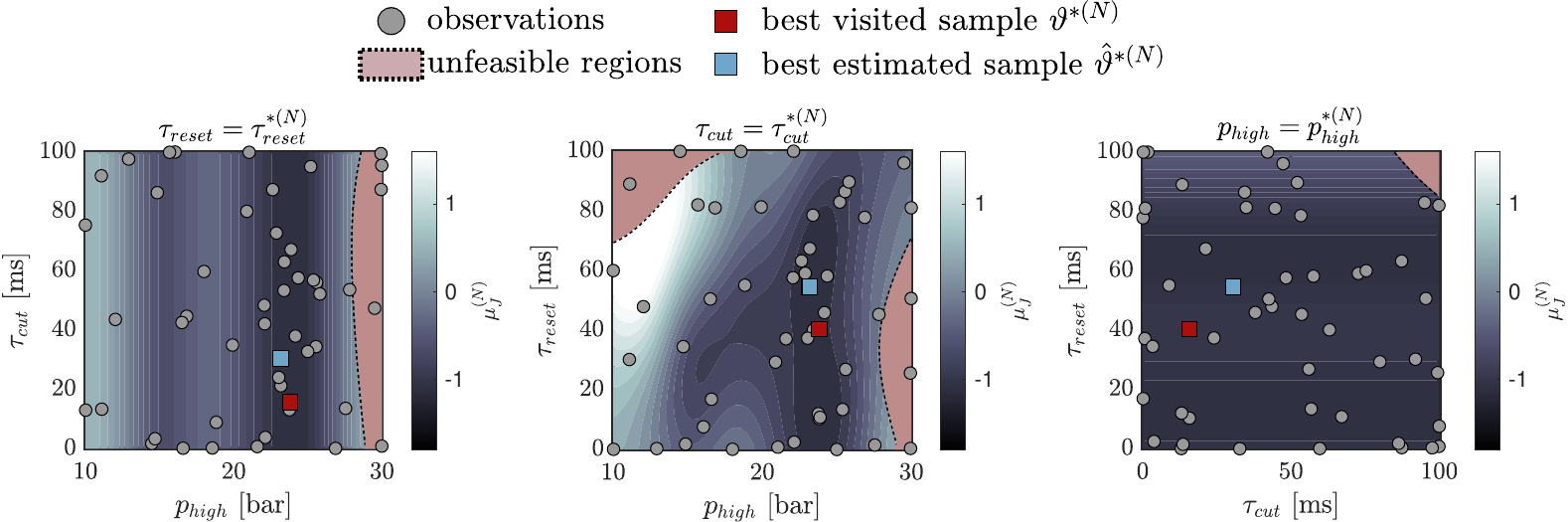}
	\caption{ \small{Samples distribution over the three-dimensional search space $ \Theta $, along with the optimal candidate and the best visited query. Shaded areas indicate infeasible regions that do not meet constraints. Each plot displays the colormap of the estimated cost function, varying two parameters, with the third held at its optimal value.} }	
	\label{fig:contourPlot_2Dsections}
\end{figure*}

Figure \ref{fig:contourPlot_2Dsections} shows the distribution of samples across the three-dimensional parameters' space. The final ($n = N$) optimal estimated candidate $\hat{\vartheta}^{*(N)}$ and the best visited query ${\vartheta}^{*(N)}$ are also depicted. Each sub-figure colormap refers to the mean value of the surrogate model for the cost function, with one parameter fixed at its optimal value; the \textit{infeasible regions} computed through to the constraint function $\hat{c}_t(\vartheta)$ are shaded. 
Although the entire search space is eventually sampled adequately to encourage \textit{exploration}, the most densely sampled area is the one associated with a higher probability of finding the minimum cost, promoting \textit{exploitation}. By observing the color gradient of the estimated cost function, it can be concluded that among the three optimization parameters the most important is the CBW step pressure amplitude $p_{high}$, followed by the reset torque delay, $\tau_{reset}$, and finally $\tau_{cut}$. The optimal value of $p_{high}$ is $23.1$ bar: lower values lead to a significant loss in gearshift smoothness due to inadequate clutch plate decoupling, while higher values, specifically beyond $26$ bar, offer minimal improvements in smoothness and result too long gearshift durations. Regarding $\tau_{reset}$, an early engine torque reset can cause engine overspeed and slightly reduce smoothness. However, it is more critical to avoid excessively delayed torque restoration, as the engine torque step would occur when the clutch is nearly closed, resulting in a notable decrease in smoothness and an extended gearshift duration. It is worth noting the low sensitivity of the parameter $\tau_{cut}$ on the gear shifting process. This is due to the trade-off between the conflicting objectives of the cost function that balance each other. Early torque cuts reduce the benefits introduced by CBW usage, favoring earlier shifting. Conversely, excessively late torque cuts take advantage of clutch opening but result in prolonged gear shifting. Overall, a slight delay promotes effective synchronization between the CBW and engine torque cut operations without significantly compromising maneuver duration. 
\begin{figure}[H]
	\centering
	\includegraphics[width=\columnwidth]{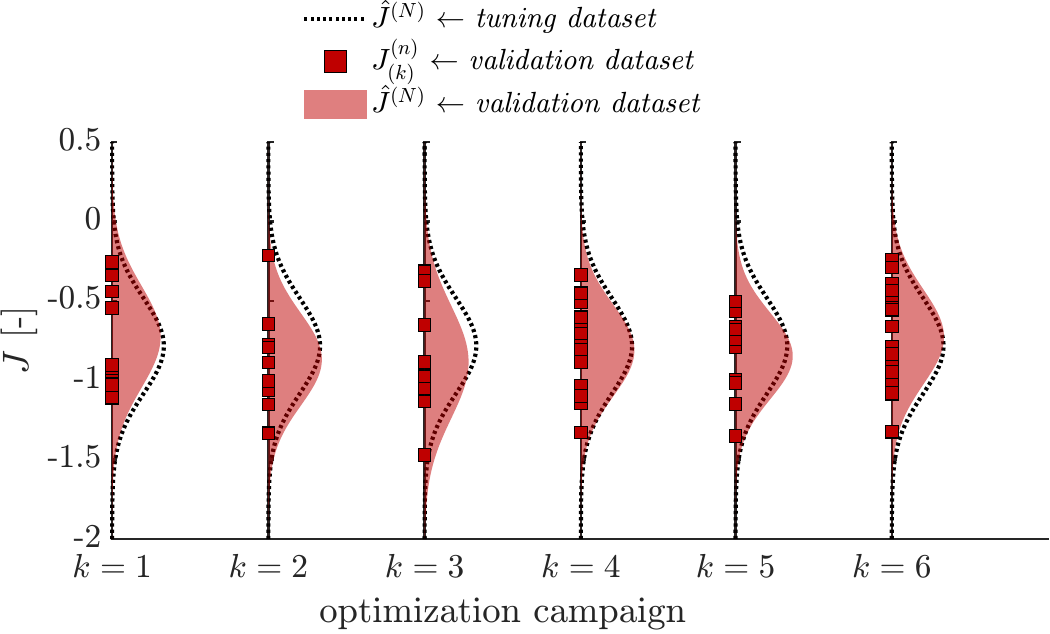}
	\caption{\small Comparison between the optimal cost function distribution estimated at the end of the optimization process, and the one estimated using the validation dataset.}
	\label{fig:Jest_vs_opt_rep}
\end{figure}
The proposed optimization campaign was repeated multiple times. In each $k$-th campaign, at the end of the optimization procedure, the retrieved optimal parameter set $\vartheta_{k}^{*(N)}$ was tested through repeated experiments to gather useful data for an \textit{a-posteriori} validation of the CBO-learned surrogate models. Figure \ref{fig:Jest_vs_opt_rep} shows, for each repetition, the estimated distribution of the cost function for the optimal parameter values, $\hat{J}^{(N)}(\vartheta_{k}^{*(N)})$ (dashed black line). This is compared with the same distribution (red shaded area), retrieved using only the validation dataset, whose samples are also represented by red squares, serving as a visual confirmation. The estimated surrogate models closely align with those obtained from the validation data, supporting the validity of the employed learning method. Moreover, the consistency of the results across different repetitions should be highlighted, reinforcing the statistical robustness of the proposed tuning approach.

\subsection{Comparison with a \textit{Random Search} optimization}
To assess the benefits of the CBO framework for parameter tuning, a comparison with a \textit{Random Search} (RS) approach is presented. The RS optimization randomly samples the paramters' search space and, similarly to the CBO optimization, selects the final optimal parameter values as the \textit{best feasible visited sample} among all the recorded values of ${J}^{(n)}$ and ${c}_i^{(n)}$.

\begin{figure*}
	\centering
	\includegraphics[width=0.77\linewidth, trim={0cm 0cm 0cm 0cm}, clip]{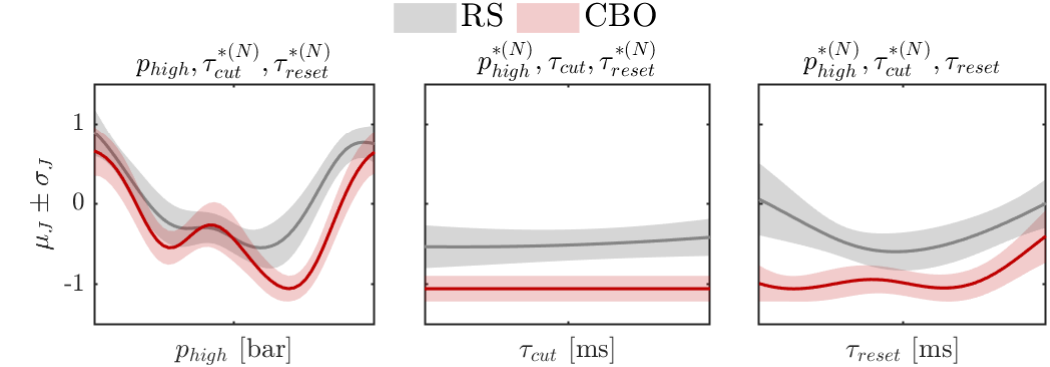}
	\caption{ \small{Mean (solid line) and standard deviation (shaded area) of the final cost function estimate \( \hat{J}^{(N)} \) as a function of its parameters. In each plot, two parameters are set to their final optimal values. CBO results are compared with those obtained from the RS approach.} }
	\label{fig:best_est_point_mean_and_std}
\end{figure*}

To make an initial comparison, the final estimated surrogate cost function model from CBO is compared with the one estimated using the collected Random Search (RS) data. Figure~\ref{fig:best_est_point_mean_and_std} shows the mean values of the two cost functions along with their associated uncertainties (related to the corresponding estimated standard deviations). Each plot represents the cost function as one parameter varies, while the other two remain constant at their optimal values. Although the main trends of the cost function are captured also using random data, the overall estimate is less precise and the higher uncertainty in model estimation eventually results in sub-optimal parameter tuning.

The \textit{convergence rate}, \textit{i.e.} the number of iterations required to identify the optimal parameter values, is addressed in the following. To mitigate the effects of process noise, we refrain from directly using the minimum cost value iteration. Instead, we estimate the convergence rate based on the similarity between the cost function estimate at each $n$-th iteration and the one learned at the end of the optimization. As long as the cost function estimate differs from the final one, the proposed optimum is unreliable. Opposite, similarity suggests the process is learned, making the optimal parameters trustworthy. To quantitatively compute these distances, we use the \textit{Kullback-Leibler} (KL) \textit{divergence metric}, following the concept of \textit{relative entropy} \cite{2008Perez}. The KL divergence quantifies the difference of a probability density function (p.d.f.) $p$ with respect to a probability density function $q$, and is defined as
\begin{equation}
	\delta^{(n)}_{k}(p^{(n)}_{k} || q_{k}) = \int_{-\infty}^{+\infty} p^{(n)}_{k}(J) \, \log \frac{p^{(n)}_{k}(J)}{q_{k}(J)} \, dJ.
\end{equation}
In the considered case,
\begin{align*}
	p^{(n)}_k(J)  \sim  \mathcal{N} \Biggl( \mu_J^{(\bm{n})} \Big( \vartheta_{k}^{* (N)} \Big) , \, \sigma_J^{(\bm{n})} \Big( \vartheta_{k}^{*(N)} \Big) \Biggl)
	\label{KL_p}
\end{align*}
corresponds to the p.d.f. of the estimated cost function at iteration $n$, evaluated in the final optimum value $\vartheta^{*(N)}$, whereas 
\begin{align*}
	q_k(J)  \sim  \mathcal{N} \Biggl( \mu_J^{(\bm{N})} \Big( \vartheta_{k}^{* (N)} \Big) , \, \sigma_J^{(\bm{N})} \Big( \vartheta_{k}^{* (N)} \Big) \Biggl)
	\label{KL_q}
\end{align*}
is the same p.d.f., at final iteration $N$.
\begin{figure}[H]
	\centering
	\includegraphics[width=\columnwidth]{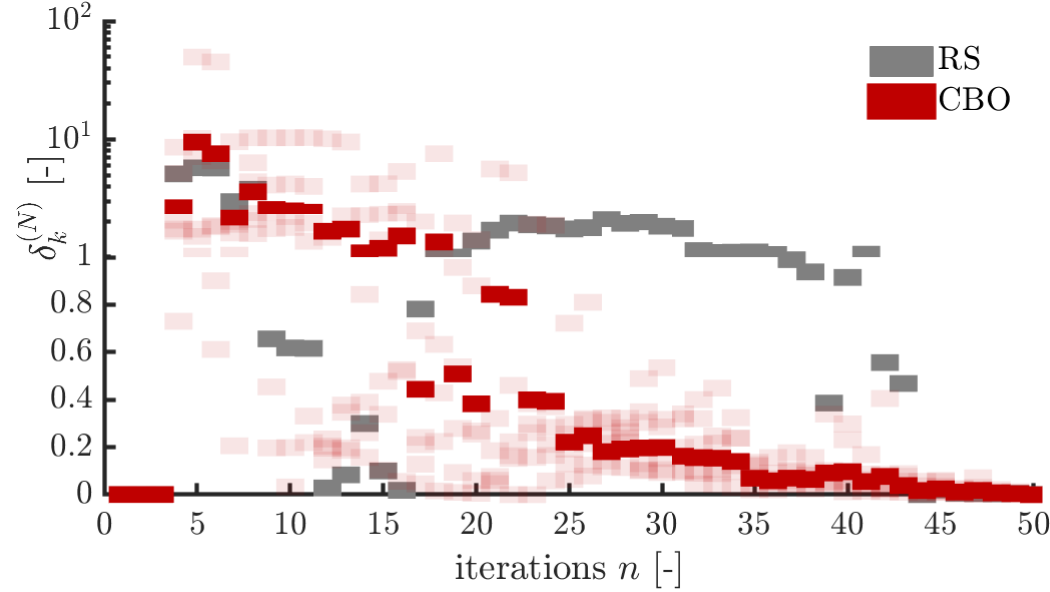}
	\caption{ \small{Evolution of the KL divergence metric. } }
	\label{fig:KL}
\end{figure}
Figure \ref{fig:KL} shows the KL metric evolution for each CBO campaign (red shaded markers) and its average (red solid markers) compared with respect to the RS process (gray marker). The first $N_r = 4$ samples are conventionally set to zero, as they correspond to the algorithm's initialization phase where no cost function estimate is available. All CBO runs demonstrate a rapid convergence to the final probability density function: only $23$ iterations are needed to reduce the average KL metric below $0.5$, indicating that the CBO p.d.f. already approximates the final one. In contrast, it takes more than $40$ iterations for the Random Search (RS) approach to converge to its final p.d.f. The reason for this behavior lies in the fact that, after the initialization, CBO actively focuses on searching for the optimal parameter regions, thereby accelerating the algorithm's convergence. Only in the second phase CBO experiments concentrate on corroborating established beliefs by sampling other underexplored regions.
\begin{figure}[H]
	\centering
	\includegraphics[scale=0.4, trim={5cm 0cm 12cm 5cm}, clip]{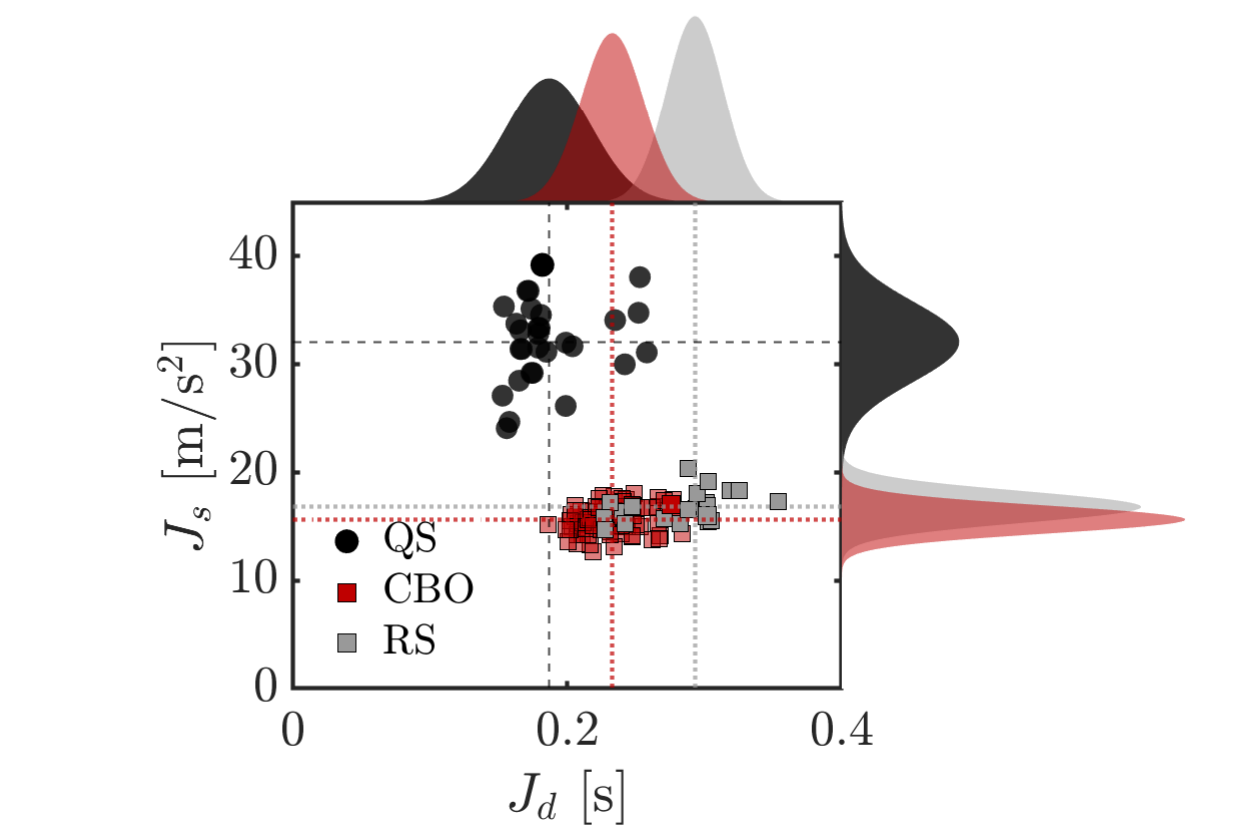}
	\caption{ \small{Smoothness and duration metrics from validation experiments of the CBO and RS campaigns, compared with samples from the QS logic. Estimated distributions are also shown.} }
	\label{fig:J_dur_rms}
\end{figure}
As a final result, we evaluate the quality of the tuning process by comparing the gearshift performances achieved with (1) \textit{QS-CBW} using the parameter values obtained through CBO optimal tuning, (2) \textit{QS-CBW} using the parameter values identified after Random Search (RS) tuning, and (3) the benchmark \textit{Quick-Shift} strategy, which does not exploit the advantages of the CBW actuator. Both smoothness and duration performance metrics are shown in Figure \ref{fig:J_dur_rms}, along with their estimated probability density functions. First, the effectiveness of the Semi-Automated Manual Transmission and the proposed \textit{QS-CBW} control is again evident. Indeed, regardless the parameters' tuning method, a significant improvement in gearshift smoothness can be achieved at the price of a slightly increased maneuver duration. Additionally, a higher maneuver consistency is achieved. As a second result, the advantage of the CBO tuning compared to a simpler RS stems out again, as the former gearshifts generally show a better smoothness index. This is due to the wiser usage of the available experiment budget done by the CBO algorithm, promoting its usage as an effective parameters tuning tool also for the considered application.

\section{Conclusions}\label{sec:conclusions}
This paper addresses the gearshift control strategy for a Semi-Automated Manual Transmission (S-AMT) powertrain layout for two-wheelers. The proposed open-loop strategy operates the engine torque reduction and the partial opening of the clutch, achieving smoother gearshifts compared to the standard \textit{Quick-Shift} strategy, with minimal compromises in terms of duration. Parameter calibration is addressed solving an optimization problem by means of the data-driven Constrained Bayesian Optimization, showing consistent results across several calibration campaign, and a faster convergence compared to a simple Random Search approach. 

The proposed framework can be generalized to different gearshifts without substantial modifications and the obtained results make the presented control strategy and the auto-calibration framework particularly attractive for industrial applications on actual vehicles. Indeed, fast real-time operations only include the computation of smoothness and duration indices, which involve simple dynamic filters and algebraic operations. While CBO may be more computationally demanding, it can be executed after each gearshift with less strict real-time requirements. Additionally, if needed, the scientific literature offers other more computational efficient data-driven global optimization algorithms, such as the one proposed in \cite{sabug2021smgo}.

\IEEEpeerreviewmaketitle
\bibliographystyle{ieeetr}
\bibliography{bibliography}

\typeout{get arXiv to do 4 passes: Label(s) may have changed. Rerun}
\end{document}